\documentclass[%
 reprint,
superscriptaddress,
 amsmath,amssymb,
 aps,
]{revtex4-2}

\usepackage{graphicx}
\usepackage{caption}
\usepackage{subcaption}
\usepackage{dcolumn}
\usepackage{bm}
\usepackage{float}
\usepackage[justification=raggedright,singlelinecheck=false]{caption}

\usepackage{acronym}
\newacro{BBH}[BBH]{Binary Black-Hole}
\newacro{PSD}[PSD]{Power Spectral Density}
\newacro{SNR}[SNR]{Signal-to-Noise Ratio}
\newacro{GW}[GW]{Gravitational Wave}
\newacro{IMBH}[IMBH]{Intermediate-Mass Black Holes}
\newacro{SBI}[SBI]{Simulation-Based Inference}
\newacro{LVK}[LVK]{LIGO-Virgo-KAGRA}
\newacro{FMPE}[FMPE]{Flow Matching Posterior Estimation}
\newacro{SNPE}[SNPE]{Sequential Neural Posterior Estimation}
\newacro{NPE}[NPE]{Neural Posterior Estimation}
\newacro{NPSE}[NPSE]{Neural Posterior Score Estimation}
\newacro{SVD}[SVD]{Singular Value Decompositions}
\newacro{LAMPE}[LAMPE]{Likelihood-free AMortized Posterior Estimation}
\newacro{JSD}[JSD]{Jensen-Shannon Divergence}
\newacro{PDF}[PDF]{Probability Density Function}

\usepackage{xcolor}

\usepackage[colorlinks=true,
            linkcolor=blue,
            citecolor=blue,
            urlcolor=blue]{hyperref}

\begin{document}

\preprint{APS/123-QED}

\title{Simulation-based Inference towards Gravitational-wave waveform systematics in  Intermediate-Mass Binary
Black Holes}%

\author{Sama Al-Shammari}
\email{al-shammaris@cardiff.ac.uk}
\affiliation{%
 School of Physics and Astronomy, Cardiff University, Cardiff, CF24 3AA, United Kingdom}%
 
\author{Alexandre G{\"o}ttel}%
\affiliation{%
Nottingham Centre of Gravity \& School of Mathematical Sciences,
University of Nottingham, University Park, Nottingham, NG7 2RD, United Kingdom}%

\author{Masaki Iwaya}
\affiliation{%
 School of Physics and Astronomy, Cardiff University, Cardiff, CF24 3AA, United Kingdom}%
\affiliation{
 Institute for Cosmic Ray Research, University of Tokyo, Kashiwanoha 5-1-5, Kashiwa, Chiba 277-8582 Japan
}
 
\author{Vivien Raymond}
\affiliation{%
 School of Physics and Astronomy, Cardiff University, Cardiff, CF24 3AA, United Kingdom}%

\collaboration{LIGO-Virgo-KAGRA Collaboration}

\date{\today}

\begin{abstract}
Parameter estimation for gravitational-wave signals is computationally demanding due to the high dimensionality of the parameter space and the cost of repeated waveform generation in traditional Bayesian inference. These analyses require $\mathcal{O}(10^8)$ likelihood evaluations and waveform generations, resulting in inference times of hours to days per event. Furthermore, discrepancies between waveform models introduce systematic uncertainties that can bias inferred source properties. To address these challenges, we propose a novel framework based on \ac{SBI} and \ac{NPE} and apply it to signals from \ac{IMBH}. In this framework, we train a single amortised neural posterior estimator on a large simulated dataset generated using two state-of-the-art waveform approximants, IMRPhenomXPHM and SEOBNRv5PHM. By treating the waveform model index as a latent variable, the network learns to produce posterior distributions that are naturally marginalized over the discrepancies of the two waveform models. Once trained, the model enables direct posterior sampling in milliseconds per event, eliminating the need for likelihood evaluations while simultaneously accounting for model systematics. 

We demonstrate that this approach recovers accurate posterior distributions for \ac{IMBH} signals injected into Gaussian noise, achieving close agreement with traditional nested-sampling results while reducing inference time by several orders of magnitude. Our results show that \ac{NPE} can robustly incorporate waveform-model systematics within a unified framework, offering a scalable path toward rapid, systematics-aware gravitational-wave inference. Establishing these methods as promising alternatives to classical likelihood-based pipelines for current and future high-mass gravitational-wave observations.

\end{abstract}

\maketitle


\section{\label{sec:intro}Introduction}
In the last decade, \ac{GW} astrophysics has grown immensely, enabling new avenues to explore various topics in cosmology, neutron star physics, and black hole astrophysics, to name a few. This is due to the rapid increase in detected \ac{GW} signals from the LIGO \cite{buikema2020sensitivity}, Virgo \cite{acernese2019increasing} and KAGRA \cite{akutsu2021overview} observatories, which together comprise the \ac{LVK} collaboration. During the third observing run, the \ac{LVK} reported over 100 \ac{GW} signals from compact binary coalscneces (CBCs) \cite{abbott2023gwtc}. The first part of the fourth observing run, which began in May 2023, has already doubled the number of total signals that were detected and published in the third observing run \cite{abac2025gwtc}. As the \ac{LVK} detectors become more sensitive and more detectors are put online in the future such as the Einstein Telescope \cite{punturo2010einstein} and the Cosmic Explorer \cite{reitze2019cosmic}, we will require enhanced methods for rapid parameter estimation that do not sacrifice on accuracy. Among the detected population, the most massive binary black hole mergers, such as GW231123 probe the poorly understood transition region between stellar-mass and intermediate-mass black holes, with component masses approaching or exceeding $\sim 100\,M_\odot$ \cite{abac2025gw231123}. Such systems are of particular astrophysical interest, as \ac{IMBH}s are thought to play an important role in stellar cluster dynamics, hierarchical black-hole growth, and galaxy evolution \cite{greene2020intermediate}. However, in the sensitive frequency band of current ground-based detectors, \ac{IMBH} mergers produce gravitational-wave signals that are extremely short in duration, often lasting only a few tens of milliseconds. This makes their parameter estimation especially challenging, as short-duration signals are more susceptible to contamination from transient noise artefacts, or glitches, which occur frequently in interferometric data \cite{abbott2023gwtc, abac2025gwtc}. As a result, \ac{IMBH} signals represent a demanding test case for inference methodologies, particularly in terms of robustness to non-Gaussian noise.

Bayesian inference is the standard framework for estimating the parameters, or source properties of the detected \ac{GW} signals. However, while traditional sampling methods \cite{skilling2004nested, ashton2022nested, speagle2020dynesty, geyer1992practical, veitch2015parameter, williams2021nested} and software packages that leverage them, such as the \textsc{Bilby} library \cite{ashton2019bilby, romero2020bayesian}, are widely used, they are computationally expensive due to the need for repeated likelihood evaluations. A single event can require $\mathcal{O}(10^{8})$ likelihood evaluations, requiring hours or days of runtime on
high-performance computing clusters, though some work has been done to accelerate likelihood evaluations \cite{cornish2010fast, cornish2021heterodyned, Vinciguerra:2017ngf, Morisaki:2021ngj, Morisaki:2023kuq, Smith:2016qas, Canizares:2014fya} as well as additional work trying to find alternative sampling algorithms to reduce the computational cost of Bayesian inference \cite{Tiwari:2023mzf, Lange:2018pyp, Fairhurst:2023idl, Nitz:2024nhj}. Additionally, while performing these traditional likelihood-based analyses, different waveform models, or templates, are utilised. These waveform models are derived using distinct approximation frameworks for the relativistic two-body problem, such as post-Newtonian expansions, effective-one-body formalisms, and numerical relativity. The use of different modelling strategies leads to multiple template families whose structural differences can result in systematic variations in the inferred source parameters. These errors and biases that can be solely attributed to the choice of waveform
models are commonly known as waveform systematics, and are the main target of our novel approach. These uncertainties are highly prevalent in areas of the parameter space where the mass ratio of the two binaries is high or when there are high spins. Furthermore, in the presence of non-Gaussian noise transients, mismatches between the data and waveform models can project differently onto the parameter space for different approximants.

To address these challenges, we adopt Simulation-based Inference \ac{SBI} methods \cite{cranmer2020frontier}, and in particular neural posterior estimation (NPE), which enables posterior inference without requiring explicit likelihood evaluations. 
This approach approximates the posterior by first generating a large, synthetic dataset of signal, sampled from the prior, then adding noise realisations. This synthetic dataset is in turn used to train a neural network to learn the mapping between observed data and the posterior distribution. This allows for amortisation of the high computational cost over future detections. There are many examples of \ac{NPE} being applied to \ac{GW} parameter estimation \cite{alvey2024simulation, bhardwaj2023sequential, dax2021real, dax2021group, dax2023flow, delaunoy2020lightning, green2020gravitational}, tackling \ac{PSD} uncertainties \cite{wildberger2023adapting}, eccentricity \cite{gupte2024evidence}, binary neutron star mergers \cite{dax2021group}, population analysis \cite{wong2020gravitational}, and our own paper addressing parameter estimation with real-detector noise \cite{raymond2024simulation}.

In this work, we utilise neural spline flows \cite{durkan2019neural} to train a normalising flow model on simulated data from \ac{IMBH}s. 
This trained network acts as a neural density estimator, which when given unseen or observed data, can sample the posterior distribution directly. The simulated data we use to train this neural density estimator was generated using two different waveform approximants with equal weights. By incorporating multiple waveform approximants, we marginalise over these systematic differences, providing a more robust posterior estimate. A similar type of analysis performed in the context of classical sampling methods can be found in \cite{ashton2020multiwaveform, hoy2022accelerating, Hoy:2024vpc}.

These types of analyses are typically performed in the frequency domain only, whereas in this work we, similar to \cite{bhardwaj2023sequential}, utilise a dual-domain approach. However, unlike their methods, here, we adopt a unified embedding approach in which both representations are provided jointly to a single \ac{NPE} inference model. Providing multiple representations of the signal can be interpreted as a form of multi-view learning, where complementary representations of the same data are incorporated jointly. This is analogous in spirit to data augmentation, although the representations are combined within a single input rather than treated as separate samples. The dual-domain representation exposes complementary structural aspects of the waveform, such as global spectral content in the frequency domain and local temporal correlations in the time domain, without introducing additional physical assumptions. By training on both domains simultaneously, the network is encouraged to learn features that are robust to representation-specific artefacts and to exploit redundancies between domains, potentially improving stability and generalisation in regions of parameter space where signals are short-lived or highly structured, such as for \ac{IMBH}s signals whose few in-band cycle signals demand efficient use of all available information. 

\section{\label{sec:methods}Methods}
The aim of this work is to infer a posterior distribution over a subset of \ac{IMBH} parameters while accounting for systematic uncertainties arising from different waveform approximants. We denote these parameters by $\theta$ and the detector data as $d$. We consider $\{w_\ell\}_{\ell=1}^{N}$, 
a discrete, finite, set of waveform models, where each model provides a prescription for computing the expected gravitational-wave signal given a set of source parameters. We denote by $H_\ell$ the hypothesis that the data were generated using waveform model $w_\ell$. Each hypothesis is assigned a prior mixing weight $\xi_\ell$, with $\sum_{\ell=1}^{N} \xi_\ell = 1$, reflecting our prior belief about the relative plausibility of each waveform model. Under these assumptions, the posterior distribution can be written as
\begin{equation}
P(\theta \mid d, { w_\ell })
\propto P(\theta)\sum_{\ell=1}^N \xi_\ell P(d \mid \theta, H_\ell).
\label{eq:mixture_post}
\end{equation} 

In standard gravitational-wave analyses assuming stationary Gaussian noise, the likelihood takes the form
\begin{equation}
P(d \mid \theta, H_\ell) \propto
\exp\!\left[-\frac{1}{2}\sum_I \langle d_I - h_I^{(\ell)}(\theta) \mid d_I - h_I^{(\ell)}(\theta) \rangle \right],
\end{equation}
where $h_I^{(\ell)}(\theta)$ is the detector response predicted by waveform model $w_\ell$ for detector $I$, and $\langle \cdot \mid \cdot \rangle$ denotes the noise-weighted inner product. Different waveform approximants therefore enter the inference problem through different model predictions for the strain $h_I^{(\ell)}(\theta)$.

Rather than explicitly evaluating and combining likelihoods for each waveform model, we adopt a likelihood-free approach. By training a single amortised neural posterior estimator on simulated data generated from all waveform models, with the model index $\ell$ treated as a latent variable drawn according to the mixing weights $\{\xi_\ell\}$, the resulting estimator implicitly learns a posterior that is marginalised over waveform-model systematics. In this work we set $N=2$, though this can be changed to include more waveform approximants in the analysis, and chose the state-of-the-art precessing approximants IMRPhenomXPHM \cite{pratten2021computationally} and SEOBNRv5PHM \cite{ramos2023next}, which are both widely used within \ac{LVK} parameter estimation in as recently as the fourth observing run and the difference in the inferred posteriors is often used as a measure of waveform systematic error \cite{abac2025gwtc}.

We define a prior over intrinsic and extrinsic source parameters sampled from astrophysically motivated prior ranges, which can be seen in Table \ref{tab:parameters}. We generate the data using the intrinsic parameters only while the extrinsic parameters are generated on-the-fly during training, such as in \cite{kofler2025flexible}. We generate each simulated signal by sampling all 15 source parameters from their prior distributions. However, the neural posterior estimator is trained to infer only a 13-dimensional subset of these parameters, corresponding to the selected training targets (or sampling keys). The orbital phase and polarisation angle are therefore varied when generating the training data but are not included in the network output. In this sense, the learned posterior is marginalised over these two parameters, which are often treated as nuisance parameters. The framework can be straightforwardly extended to infer them as well by including them among the training targets.

\begin{table*}[ht]
    \centering
    \begin{tabular}{l c c}
        \textbf{Parameter} & \textbf{Range} & \textbf{Prior distribution} \\
        \hline
        Geocentric time $t_c$ 
            & $1126259632.403$ -- $1126259632.423$ s 
            & Uniform \\

        Mass ratio $q$
            & $0.3$ -- $1.0$ 
            & Uniform \\

        Chirp mass $\mathcal{M}$
            & $80$ -- $120~M_\odot$
            & Uniform \\

        Luminosity distance $d_L$
            & $100$ -- $5000$ Mpc
            & LogUniform \\

        Right ascension $\mathrm{RA}$
            & $0$ -- $2\pi$
            & Uniform (periodic) \\

        Declination $\mathrm{DEC}$
            & $-\pi/2$ -- $+\pi/2$
            & Cosine \\

        Inclination angle $\theta_{JN}$
            & $0$ -- $\pi$
            & Sine \\

        Polarisation angle $\psi$
            & $0$ -- $\pi$
            & Uniform (periodic) \\

        Coalescence phase $\phi_c$
            & $0$ -- $2\pi$
            & Uniform (periodic) \\

        Spin magnitude $a_1$
            & $0$ -- $0.99$
            & Uniform \\

        Spin magnitude $a_2$
            & $0$ -- $0.99$
            & Uniform \\

        Spin tilt angle $\theta_1$
            & $0$ -- $\pi$
            & Sine \\

        Spin tilt angle $\theta_2$
            & $0$ -- $\pi$
            & Sine \\

        Spin azimuthal difference $\Delta\phi$
            & $0$ -- $2\pi$
            & Uniform (periodic) \\

        Azimuth of orbital angular momentum around total angular momentum $\phi_{JL}$
            & $0$ -- $2\pi$
            & Uniform (periodic) \\
        \hline
    \end{tabular}
    \caption{Prior distributions of the parameters used to generate simulated gravitational-wave signals. These parameters define the training prior for the \ac{SBI} neural networks; parameters not included in the network output are treated as latent variables and marginalised over during training. The coalescence time is defined relative to the injection time within each simulated segment, rather than as an absolute GPS time, and is therefore arbitrary up to a constant offset.}
    \label{tab:parameters}
\end{table*}

We sample from the prior and compute the \ac{GW} frequency-domain polarisations $h_+(f,\theta), \quad h_\times(f,\theta)$ for each waveform model according to
\begin{equation}
\{h_+(f),h_\times(f)\}_\ell = w_\ell(\theta; f).
\end{equation}

We compute the antenna pattern factors $F_{+,I}$ and $F_{\times,I}$ for each interferometer $I$. The frequency-domain strain for the $I^{th}$ detector is
\begin{equation}
d_I(f,\theta)
    = F_{+,I}\,h_+(f,\theta)
    + F_{\times,I}\,h_\times(f,\theta),
\end{equation}
with appropriate time-shift factors to account for differences in light-travel time between detectors. The strain is truncated to a short, fixed-duration segment surrounding the merger, in this work this was $0.5$ s for each simulated signal due to the large masses of the mergers used for the analysis. The strain is whitened using a fixed, reference one-sided noise power spectral density $S_n(f)$ corresponding to the Advanced LIGO design sensitivity. This PSD is not estimated independently for each segment, but instead applied uniformly across all simulations. Gaussian noise realisations are then generated consistently with this PSD and added to the signal. As a result, the network is exposed to noise through repeated realisations during training, rather than through explicit PSD estimation, allowing it to learn the noise structure implicitly. While we do utilise Gaussian noise in this work, this method could be extended to use real detector noise.

Although whitening and noise generation occur in the frequency domain, the time-domain structure remains important, particularly for high-mass, precessing binaries, where the limited number of observable cycles compresses complex spin–orbital dynamics into a narrow time window, increasing sensitivity to waveform approximant differences and noise realisations \cite{miller2025measuring}. We therefore compute the whitened time-domain strain via inverse Fourier transform, and for each detector we retain (i) the real and imaginary components of the whitened frequency-domain strain, and (ii) the whitened time-domain strain. These components are concatenated across detectors into a 3-dimensional vector encoding both spectral and temporal information.

The concatenated time-frequency representation of the whitened detector strain typically contains tens of thousands of dimensions once the real and imaginary frequency components and the corresponding time-domain samples are combined across detectors. We employ a nonlinear embedding network that compresses this high-dimensional strain vector into a lower-dimensional representation optimised jointly with the inference model.

The embedding network is a deep multilayer perceptron comprising twenty fully connected layers. 
It begins with four wide layers of 512 neurons, followed by four layers of 256 neurons, four of 128 neurons, four of 64 neurons, and finally four compact layers of 32 neurons, each followed by an ELU activation function. This progressively narrowing architecture acts as a learned nonlinear compression of the high-dimensional time-frequency strain input. Because the embedding is trained jointly with the conditional density estimator, it automatically learns which spectral and temporal features are most informative for parameter estimation. This allows the model to preserve subtle physical structure such as waveform–model discrepancies without relying on handcrafted or linear compression schemes. The training was done using $10^{6}$ simulated data segments from each approximant , with time and frequency domain signals for each injection concatenated as the input data, on a A100 GPU. Training typically required between 5 and 7 days to converge, depending on the early-stopping criterion.

The resulting latent vector $z$ serves as the conditioning input to a \ac{NPE} model based on conditional normalising flows. We employ neural spline flows as implemented in the \textsc{sbi} framework \cite{BoeltsDeistler_sbi_2025}, which model the posterior density $q_\phi(\theta \mid z)$ through a sequence of invertible transformations that map samples from a simple base distribution (typically a standard Gaussian) to the target posterior. Conditioning on $z$ allows the flow to adapt its transformation to the observed strain data.

Training proceeds by maximising the conditional log-likelihood of the parameters under the flow model,
\begin{equation}
\mathcal{L}(\phi)
= -\mathbb{E}_{(\theta, z)} \left[ \log q_\phi(\theta \mid z) \right],
\end{equation}
where the expectation is approximated using batches of simulated data. 
Optimisation is performed using the AdamW optimiser with a learning rate of $10^{-4}$, 
cosine-annealing learning-rate scheduling, and gradient clipping to stabilise training. 
A summary of the training hyperparameters is provided in Table~\ref{tab:training}. 
The normalising flow architecture consists of multiple spline-based transformations with residual connections, 
allowing the model to represent highly non-Gaussian, multimodal posteriors characteristic of gravitational-wave inference.

\begin{table}[ht]
    \centering
    \begin{tabular}{l c}
        \textbf{Hyperparameter} & \textbf{Value} \\
        \hline
        Optimiser & AdamW \\
        Learning rate & $1 \times 10^{-4}$ \\
        Scheduler & CosineAnnealingLR \\
        $T_{\max}$ & 1024 \\
        Gradient clipping & 5.0 \\
        Batch size & 8192 \\
        Number of training samples & $2 \times 10^{6}$ \\
        Number of transforms & 64 \\
        Hidden features & 256 \\
        \hline
    \end{tabular}
    \caption{Training hyperparameters used for optimisation of the neural posterior estimator. The dataset consists of $10^{6}$ samples generated from each waveform approximant.}
    \label{tab:training}
\end{table}

Once trained, the NPE network provides an amortised approximation to the posterior distribution. For any observed or simulated dataset $d_{\mathrm{obs}}$, the trained normalising flow then yields the conditional posterior $q_\phi(\theta \mid d_{\mathrm{obs}})$, from which samples can be drawn quickly. This enables fast, likelihood-free parameter estimation that marginalises over waveform-model systematics, nuisance parameters, and Gaussian noise realisations, reducing inference time to milliseconds.

\section{\label{sec:results}Results}

The results outlined in this section were obtained using \ac{SBI} networks trained on simulated data from spinning \ac{IMBH} systems. The training data were generated in a 15-dimensional parameter space, with signals injected into Gaussian noise in the frequency domain and concatenated with their inverse Fourier transforms in the time domain, for both LIGO detectors. Of these parameters, 13 are inferred by the network, while the coalescence phase and polarisation angle are treated as latent variables and are therefore marginalised over during training. Half of the training data were generated using the IMRPhenomXPHM waveform model, and the other half using SEOBNRv5PHM, corresponding to an equal prior weighting over the two approximants. This construction allows the network to learn posterior distributions that are marginalised over both extrinsic nuisance parameters and waveform modelling systematics.

\begin{figure*}[!t]
    \centering
    \includegraphics[width=\linewidth]{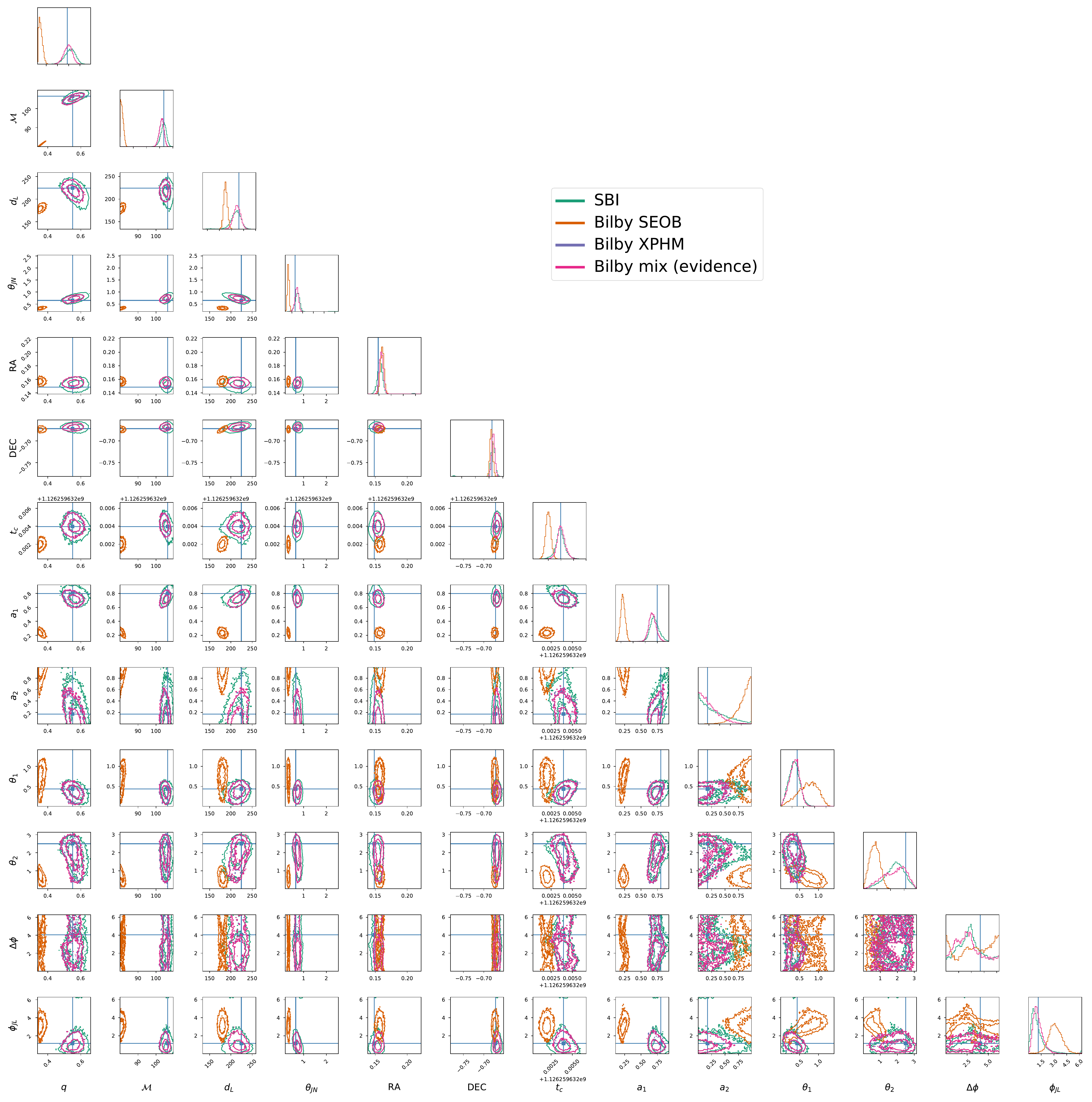}
    \caption{Example posterior recovery for an \ac{IMBH} injection of type IMRPhenomXPHM in Gaussian noise. We compare the waveform-marginalised \ac{NPE} posterior (trained jointly on IMRPhenomXPHM and SEOBNRv5PHM) against evidence-weighted posteriors obtained from separate \textsc{bilby}--\textsc{dynesty} analyses using each waveform approximant. The agreement demonstrates that the amortised \ac{NPE} network reproduces the combined waveform-marginalised inference while reducing inference time by orders of magnitude. The results obtained from the \textsc{bilby}--\textsc{dynesty} analysis using IMRPhenomXPHM and the evidence-weighted posterior are indistinuishable throughout the plot due to the low evidence value from the SEOBNRv5PHM analysis.}
    \label{fig:posterior}
\end{figure*}

\begin{figure}[t]
    \centering
    \includegraphics[width=\linewidth]{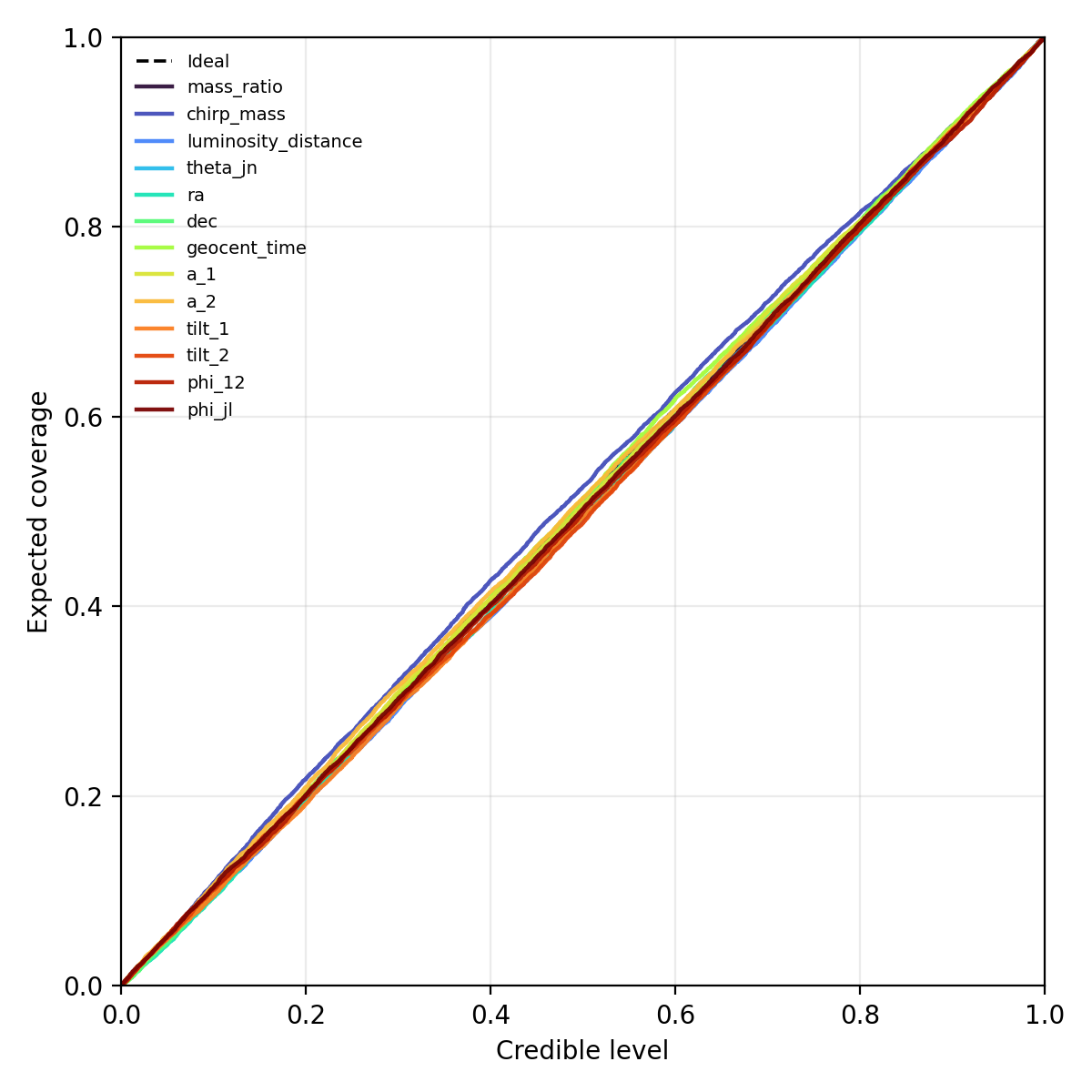}
    \caption{Percentile--percentile (PP) test for the waveform-marginalised \ac{NPE} network using $10^4$ synthetic injections ($5000$ per waveform family). The diagonal corresponds to perfectly calibrated credible intervals, shaded bands indicate the expected statistical fluctuations under ideal calibration. The close agreement with the diagonal indicates well-calibrated posterior coverage across the 13 inferred parameters.}
    \label{fig:ppplot}
\end{figure}

The \ac{SBI} network is expected to converge towards a representation of the conditional posterior mapping, enabling direct sampling from the posterior with the corresponding estimated gaussian likelihood. In order to check how well the training has converged, we generate a set of IMRPhenomXPHM data injected into Gaussian noise and perform parameter estimation with both \ac{SBI} and \textsc{bilby}--\textsc{dynesty} \cite{speagle2020dynesty}. We use the same Advanced LIGO zero-detuned high power noise curve as in the training data generation. In the \ac{SBI} case, we recover the injection using the novel network we propose in this project. The Dynesty recovery per injection was also performed with both waveform approximants separately and then an extra distribution was generated using both of the resultant posterior samples from each analysis and recovery, weighted by their evidence following Bayesian methods such as \cite{ashton2020multiwaveform, hoy2022accelerating}, to make a new distribution that combines the information of both distributions and marginalises over systematics and uncertainties. An example is shown in Figure~\ref{fig:posterior}, where we compare the posterior obtained from the waveform-marginalised \ac{NPE} network (‘SBI’) with posteriors recovered using \textsc{bilby}-\textsc{dynesty} under the IMRPhenomXPHM (‘Bilby XPHM’) and SEOBNRv5PHM (‘Bilby SEOB’) waveform models, as well as their evidence-weighted combination (‘Bilby mix’). The strong agreement between the \ac{SBI} and the evidence weighted combination distributions indicates that the \ac{NPE} network accurately captures the waveform-marginalised posterior. This agreement can be made better by performing a hyperparameter optimisation search to further fine-tune the learned network and make for a better and more efficient convergence of the normalising flow. The low evidence value, and thus poor agreement in the SEOBNRv5PHM distribution, is due to this particular injection being an IMRPhenomXPHM one with a high chirp mass, highly spinning primary object and a mass ratio towards the upper bound of the prior range which is known to exacerbate discrepancies between waveform approximants. Similar results were seen from all of the ten injections used for the inference stage. It is worth noting that the results obtained using \textsc{bilby}-\textsc{dynesty} under the IMRPhenomXPHM waveform model and the results obtained using the evidence-weighted mixture are indistinguishable throughout the figure due to the low evidece value from the SEOBNNRv5PHM analysis in this case. Additionally, we perform a 13-dimensional percentile-percentile test on $10^{4}$ injections, half of which were IMRPHenomXPHM injections and the other half were SEOBNRv5PHM injections, recovered with our waveform-marginalisation network. The large number of injections, enabled by the millisecond-level inference of the amortised model, permits a statistically meaningful calibration test far beyond what is typically feasible with traditional sampling-based analyses, and demonstrates that the credible intervals behave as expected, the result of this test is visualised in Figure \ref{fig:ppplot} \cite{cook2006validation}.

\section{\label{sec:discussion}Discussion}

The results of this work demonstrate that simulation-based inference, and in particular the \ac{NPE} neural spline flow framework, can provide a powerful and computationally efficient alternative to traditional likelihood-based parameter estimation coupled with waveform systematic marginalisation for intermediate-mass binary black holes. Once trained, our neural network produces samples directly from the conditional posterior distribution $q(\theta\mid d)$ in a matter of milliseconds. This represents a reduction in inference time by several orders of magnitude when compared with nested sampling via \textsc{bilby}-\textsc{dynesty}, which typically requires hours to days of computation per event.

A key achievement of this study is the simultaneous incorporation of sources of systematic variation in the differences between waveform approximants. By training on simulated data generated from two state-of-the-art precessing waveform families (IMRPhenomXPHM and SEOBNRv5PHM), the network learns a posterior that is automatically marginalised over waveform-model uncertainty. Likewise, by providing the inference model with both whitened frequency-domain data (real and imaginary components) and the corresponding whitened time-domain strain, we exploit complementary spectral and temporal information. Together, these design choices allow our \ac{NPE} model to retain astrophysically relevant features, capture waveform–model discrepancies, and achieve robust performance across a systematics-prone region of the parameter space.

The resulting posterior estimates demonstrate agreement with traditional Bayesian analyses while offering dramatically reduced computational cost. Our findings show that likelihood-free methods can be used not only to accelerate inference, but also to coherently combine information from multiple waveform families, which is a task that would traditionally require running separate stochastic sampling analyses and combining their posteriors through evidence-weighting. In contrast, \ac{NPE} learns this marginalisation directly from simulation, providing an end-to-end framework for systematics-informed gravitational-wave inference.

Several limitations remain. First, our analysis does not incorporate real detector noise or explicitly model non-Gaussian transient glitches, although the framework is readily extendable to such noise distributions given appropriate simulations. Second, while our analysis includes spin magnitudes and tilt angles and employs fully precessing waveform approximants, we do not attempt a systematic exploration of the full space of precessional dynamics, such as extreme or transitional precession regimes, nor do we investigate additional physical effects that can couple to spin precession, including orbital eccentricity or higher-order multipoles beyond those included in the adopted waveform models. Third, numerical-relativity surrogate models were not employed due to their limited coverage of the high-mass parameter space considered here, extending surrogate models or combining them with phenomenological or effective-one-body approximants would broaden the scientific reach of waveform-marginalised SBI. Finally, although the fully learned embedding network performs well, optimising its architecture, along with exploring alternative compression schemes such as invertible neural networks or convolutional encoders or other architectures is a promising direction for future work. A deeper study of hyperparameter sensitivity, data efficiency, and training stability will further strengthen the viability of \ac{SBI} for routine gravitational-wave parameter estimation. Work addressing these extensions, including the incorporation of more realistic noise and a broader exploration of precessional and dynamical effects, is currently in progress.

Overall, this work establishes a foundation for systematics-aware, dual-domain, waveform-marginalised simulation-based inference in gravitational-wave astronomy. With continued development, \ac{NPE}-based architectures have the potential to become a fast, flexible, and robust alternative to traditional parameter-estimation pipelines, particularly in regions of parameter space where waveform uncertainties are most significant.

\begin{acknowledgments}
This work was supported by the UKRI CDT in Artificial Intelligence, Machine Learning and Advanced Computing, the UK Science and Technology Facilities Council grant ST/V005618/1 and UKRI2489, the Royal Society Award ICA\textbackslash R1\textbackslash 231114 and the Leverhulme Trust Fellowship IF-2024-038.
This research has made use of data or software obtained from the Gravitational Wave Open Science Center (gwosc.org), a service of the LIGO Scientific Collaboration, the Virgo Collaboration, and KAGRA. This material is based upon work supported by NSF's LIGO Laboratory which is a major facility fully funded by the National Science Foundation, as well as the Science and Technology Facilities Council (STFC) of the United Kingdom, the Max-Planck-Society (MPS), and the State of Niedersachsen/Germany for support of the construction of Advanced LIGO and construction and operation of the GEO600 detector. Additional support for Advanced LIGO was provided by the Australian Research Council. Virgo is funded, through the European Gravitational Observatory (EGO), by the French Centre National de Recherche Scientifique (CNRS), the Italian Istituto Nazionale di Fisica Nucleare (INFN) and the Dutch Nikhef, with contributions by institutions from Belgium, Germany, Greece, Hungary, Ireland, Japan, Monaco, Poland, Portugal, Spain. KAGRA is supported by Ministry of Education, Culture, Sports, Science and Technology (MEXT), Japan Society for the Promotion of Science (JSPS) in Japan; National Research Foundation (NRF) and Ministry of Science and ICT (MSIT) in Korea; Academia Sinica (AS) and National Science and Technology Council (NSTC) in Taiwan.
The authors are grateful for computational resources provided by the LIGO Laboratory and Cardiff University and supported by National Science Foundation Grants PHY-0757058 and PHY-0823459, and STFC grants ST/I006285/1 and ST/V005618/1.
\end{acknowledgments}

\appendix

\nocite{*}

\bibliography{paper}

\end{document}